\documentclass[superscriptaddress,showpacs,twocolumn,aps,pre,10pt]{revtex4-1}
\usepackage{natbib}
\usepackage{graphicx,dcolumn,bm,bbm,amsmath,amssymb,color}

\newcommand{\eeq}{ \end{equation} }
\newcommand{\beq}{ \begin{equation} }
\newcommand{\bea}{\begin{eqnarray}}
\newcommand{\eea}{\end{eqnarray}}

\begin{document}
\title{Pickering emulsions stabilized by oppositely charged colloids: stability and pattern formation}
 
\author{Sam David Christdoss Pushpam, Madivala G. Basavaraj, and Ethayaraja Mani}
\email{ethaya@iitm.ac.in}
\affiliation{Polymer Engineering and Colloid Science Lab, Department of Chemical Engineering, Indian Institute of Technology Madras, Chennai - 600036, India}
 
\date{\today}

\begin{abstract}
Binary mixture of oppositely charged of colloids can be used to stabilize water-in-oil or oil-in-water emulsions. A Monte Carlo simulation study to address the effect of charge ratio of colloids on the stability of Pickering emulsions is presented. The colloidal particles at the interface are modeled as aligned dipolar hard spheres, with attractive interactions between unlike-charged and repulsive interaction between like-charged particles. The optimum composition (fraction of positively charged particles) required for the stabilization corresponds to a minimum in the interaction energy per particle. In addition, for each charge ratio, there is a range of compositions where emulsions can be stabilized. The structural arrangement of particles or the pattern formation at the emulsion interface is strongly influenced by the charge ratio. We find well-mixed isotropic, square and hexagonal arrangement of particles on emulsion surface for different compositions at a given charge ratio. Distribution of coordination numbers is calculated to characterize structural features. The simulation study is useful for rational design of Pickering emulsifications wherein oppositely charged colloids are used, and for the control of pattern formation that can be useful for the synthesis of colloidosomes and porous-shells derived from thereof.
\end{abstract}
 
\maketitle
\section{Introduction}
Colloids and nanoparticles readily adsorb at fluid-fluid interfaces leading to stable emulsion droplets and bubbles \cite{nallamilli,nallamilli2015,amalvy,lin,leunissen,bleibel}. Though known for long-time, the approach has been recently extended to use colloids of several kind such as polymeric latex \cite{amalvy,binks,witten,zhulina}, soft-colloids \cite{sun,herzig}, anisotropic colloids \cite{dugyala}, proteins \cite{damodaran,clark}, Janus colloids \cite{walther,faria} and patchy colloids \cite{pawar}. The recent developments in the synthesis of these new types of colloids have broadened the applications ranging from the preparation of stable emulsions and bubbles with extended period of stability  to the making of colloidosomes \cite{lou,dinsmore,easwaramoorthi,liu}. 
We recently reported experimental studies demonstrating the advantages of using oppositely charged colloids to stabilize water-in-oil (W/O) emulsions \cite{nallamilli,nallamilli2015}. It was shown that highly charged particles alone cannot stabilize emulsions because of image charge-effect at the oil-water interface that prevents adsorption of particles \cite{binks2008}. Additives such as salt or oppositely charged surfactant or oppositely charged particles are needed to overcome the repulsive barrier and to enhance the adsorption of particles to the interface. \\
In using oppositely charged particles, the key is to first form aggregates of oppositely charged particles in water and then emulsify with oil, wherein the aggregates are transported to the interface as these aggregates have low net charge. In such emulsion systems, the stability and size-distribution of the emulsion droplets strongly depend on several factors such as charge ratio, number ratio, concentration of particles and contact angles of the particles \cite{nallamilli2015}. \\

In a conventional emulsification process, tiny water droplets are first created in oil by providing mechanical energy externally. Therefore, the initial size-distribution of the drops depends on the mixing process. If the adsorption of colloids at the interface is favoured and instantaneous, and if sufficient number of colloids are available for the newly created oil-water interface, all the drops will be stable against further coalescence as the presence of particles at the interface offers steric stability. However, in practice, water drops formed initially are polydisperse and only a few of the particles or aggregates that overcome image-charge effect can adsorb at the interface. This leads to partial coverage and further coalescence. In such cases, the final drop size-distribution is governed by drop coalescence. Essentially, the drop size distribution is governed by the competition between the rates of adsorption of colloids versus rate of coalescence of the drops \cite{whitesides,arditty}. \\

When oppositely charged colloids are used, the ability of particles to adsorb on the W/O interface depends on mutual interactions between the particles. For an emulsion drops to be stable, sufficient packing of particles at the interface is necessary to prevent coalescence leading to phase separation. In the present work, we ask the following questions relevant to experimental success in making stable emulsions: what is the range of number ratio of oppositely charged colloids for a given charge density ratio that yield stable emulsion? How does this range of number ratio changes when charge density ratio is varied? What are the structural features of particles assembled at emulsion surface? To answer these questions, we consider an isolated emulsion droplet with a particle coverage of 74\%, and calculate the minimum energy per particle for a given number ratio and charge ratio using Monte Carlo simulations. In this approach, one can avoid the detailed kinetic mechanisms of adsorption of particles to the interface, which are limited by diffusion and adsorption kinetics. Thereby, we link stability of emulsions to inter-particle interactions, as the later can be tuned experimentally. \\

Another avenue for the importance of Pickering emulsion is that they have been increasingly used as a template for the synthesis of colloidosomes. With a binary mixture of particles as in the present case, such emulsion drops can be used to make colloidosomes with multi-functionality. Further, by selective removal of one of the types of particles, pores of selective size can be made, leading to application in controlled drug delivery using colloids. Such porous colloidosomes have potential use in size and shape selective transport. To this end, it is imperative to study the structural pattering of oppositely charged particles on emulsion surface and how it is influenced by charge ratio and number ratio. We address these aspects from our simulations. Structural features on colloidosomes stabilized by polarizable hard spheres were shown to be dependent of particle density \cite{fantoni}. On flat two dimensional interfaces consisting of oppositely charged surfactant species, the structural transition from lamellae to hexagonal ordering has been reported \cite{loverde2006,loverde2007}. Rich phase behaviour of binary mixtures of uncharged particles with different size ratios and compositions to study random metal alloys and metallic glasses have been reported \cite{miracle,ohkubo,miracle2003,sadoc,sadoc1973,polk}. Binary mixtures of like-charged colloids \cite{assoud2010} and dipolar colloids \cite{assoud} have been shown to exhibit diverse variety of crystalline phases in two-dimensions at zero temperature. The role of inter-particle interactions on the stability of colloidosomes has been studied for colloids with competing interactions \cite{mani2010}. In this context, our work addresses the role of inter-particle interactions on the pattern formation of oppositely charged colloids at emulsion surfaces. These results can be of use in screening emulsions that mimic the target structure of desired colloidosomes. 
\section{Model and Simulation}
\subsection{Effective interaction}
\begin{figure} 
\centering
\includegraphics [scale=0.45]{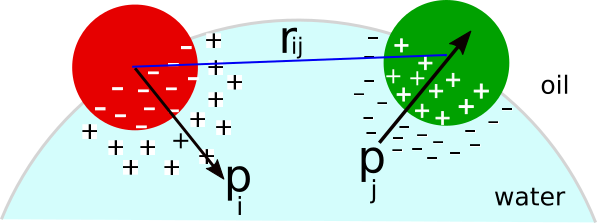}
\caption{ (Colour online) A schematic of oppositely charged colloids at the water-in-oil emulsion drop interface.}
\label {fig:dipole}
\end{figure}

In the formulation of the problem, an isolated water-in-oil emulsion droplet of radius $R$ with a total of $N_t$ number of oppositely charged colloids adsorbed on its surface is considered. The role of water, oil and oil-water interface is considered via an effective interaction potential between oppositely charged colloids at the interface. It is known that charged particles at the interface behave like dipoles because of the asymmetric distribution of charges across the interface \cite{pieranski,horozov,law,frydel}. The chargeable groups on the part of the particle immersed dissociate. Due to this charge asymmetry about the oil-water interface, the particle acts as an electric dipole (see Fig.\ref{fig:dipole}). As the system consists of oppositely charged colloids, there are attractive dipolar interactions between unlike-charged colloids and repulsion between like-charged colloids. The strength of interactions depend on the charge density, dielectric constant of water and Debye screening length. We use the pair potential of the form that depend on the relative orientation of the dipoles as
\begin{equation} \label {eq:pot}
\begin{aligned}
U(r_{ij}) = 
\begin{cases}
\infty, &  \text{if} \quad r_{ij} < \sigma  \\ 
\frac {{p_i}.{p_j} - 3(\hat{r}_{ij}.{p_i})(\hat{r}_{ij}.{p_j})}{4 \pi \epsilon \epsilon_0 r_{ij}^3}, & \text{otherwise}
\end{cases}
\end{aligned} 
\end{equation} 

where ${p_i}$ and ${p_j}$ are the electric dipole moment of particles $i$ and $j$, respectively, $\epsilon$ is the dielectric constant of water (where particles were initially present), $\epsilon_0$ is the permittivity of vacuum, $r_{ij}$ is the inter-particle separation distance, and $\hat{r}_{ij}$ is the unit vector along the vector connecting the two particles. For $r_{ij} < \sigma$, diameter of the particle, we use hard-sphere potential. The dipole moments can be related to the number of charges as ${p_i}=z_i q_i e \hat{u}_i/\kappa$ \cite{pieranski}. The parameter $1/\kappa$ is Debye screening length, $e$ is elementary charge, $q_i$ is the number of charges per particle, $z_i$ is the sign of the charge ($z_i = -1$ for negatively charged particles and $z_j = +1$ for positively charged particles), and $\hat{u}_i$ is the unit vector in the direction of the dipole moment of particle $i$.  It is the outward unit normal drawn from the center of mass of particle $i$. All the particles are constrained such that the center of mass of particles lies on the spherical surface.
Eq. \eqref{eq:pot} can now be written as
\begin{equation} \label {eq:pot2}
\begin{aligned}
U(r_{ij}) = \frac {z_i z_j q_i q_j e^2 [\hat{u}_i.\hat{u}_j - 3(\hat{n}.\hat{u}_i)(\hat{n}.\hat{u}_j)]}{4 \pi \epsilon \epsilon_0 \kappa^2 r_{ij}^3}
\end{aligned}
\end{equation}

In the simulation, energy is scaled in the units of $k_BT$ ($k_B$ is Boltzmann constant and T is temperature), and length is scaled by $\sigma$. The scaled potential then becomes
\begin{equation} \label {eq:potnondim}
\begin{aligned}
\bar{U}(\bar{r}_{ij}) = \frac {A z_i z_j q_i q_j e^2 [\hat{u}_i.\hat{u}_j - 3(\hat{n}.\hat{u}_i)(\hat{n}.\hat{u}_j)]}{\bar{r}_{ij}^3}
\end{aligned}
\end{equation}
where the prefactor $A$ is defined as

\begin{equation} \label {eq:parA}
\begin{aligned}
A = \frac{\lambda_B}{\bar{\kappa}^2\sigma}
\end{aligned}
\end{equation}
The parameter $\lambda_B$, Bjerrum length, is defined as 
\begin{equation} \label {eq:bjerrum}
\begin{aligned}
\lambda_B = \frac{e^2}{4\pi\epsilon\epsilon_0 k_BT}
\end{aligned}
\end{equation}
Debye screening length, $\kappa^{-1}$ is defined as,
\begin{equation} \label {eq:kappa}
\begin{aligned}
\kappa^{-1} = \sqrt{\frac{\epsilon\epsilon_0 k_BT}{2N_Ae^2I}}
\end{aligned}
\end{equation}
In this equation, $N_A$ is the Avogadro number and $I$ is the ionic strength due to counterions released to water depending upon the number ratio of oppositely charged particles. Note that the value of the prefactor $A$ changes with Debye screening length, which in turn changes as the composition of charged colloids is varied. To generate initial positions (coordinates) of the colloids on a spherical surface, we create $N_t$ particles with diameter of unity on a spherical surface of radius $\bar{R}$  using icosadeltahedral configuration \cite{saff}. Of the total $N_t$  particles, $N_+$ number of positively charged and $N_-$ number of negatively charged colloids of equal size. They are randomly positioned on the surface of the sphere. We fix $N_t$ = 1092 particles and study total area coverage ($\phi$) of 74\% corresponding to the radius of the sphere ($\bar{R}$) of 9.6 (assuming the particles at the oil-water interface makes a contact angle of 90$^{\circ}$). $N_t$, $\bar{R}$,   and $\phi$ are related by the geometric relation

\begin{equation} \label {eq:fraction}
\begin{aligned}
\phi = \frac{N_t a}{S} = \frac{N_t}{16\bar{R}^2}
\end{aligned}
\end{equation}

where $a$ is the cross sectional area of the particles at the interface and $S$ is the surface area of emulsion drop. 
\subsection{Monte Carlo simulation}
In the Monte Carlo simulation, a particle is randomly chosen and moved with a constraint that it remain on the spherical surface after the move. This step mimics the motion of colloids at the water-oil interface. The total interaction energy, defined as the sum of all pair-wise interactions, is calculated before ($E_o$) and after the move ($E_n$). Because the potential is long-ranged [Eq. \eqref{eq:potnondim}], in calculating total energy, each particle interacts with all the other particles. If $E_n < E_o$, the new position is accepted and continued, otherwise the new position is accepted with a probability of $e^{-(E_n - E_o)/k_BT}$. The simulation is continued until the total energy converges to equilibrium value. To speed up the equilibration of the system, we also perform swap moves along with local moves using long-range Kawasaki exchange method \cite{tamayo}, wherein two particles are randomly chosen and their positions are interchanged. The exchange is accepted with a probability:
\begin{equation}
\begin{aligned}
P = \frac{e^{-(\Delta E/k_BT)}}{1+e^{-(\Delta E/k_BT)}}
\end{aligned}
\end{equation}
Swap moves are done only during equilibration cycle with a probability of 0.1.

Constant number-volume-temperature (NVT) MC simulation is carried out for 21 different charge ratios in the range of 1 to 5 with an interval of 0.2. Charge ratio is defined as $R_c=q_-/q_+$.  $q_+$ is fixed at 2000. $q_-$ is varied according to the charge ratio. Table 1 shows the values of parameter $A$ for different compositions ratios at $R_c$ = 1. Similarly the parameter $A$ is calculated for other charge ratios from Eq. \eqref{eq:parA}. Composition of the particles on the spherical surface is expressed as the number fraction  $\phi_+ = N_+/(N_+ + N_-)$. Compositions in the range of 15 - 98\% are studied for each of the charge ratio. The simulation box size is kept at 40$\sigma$. Interaction between emulsion drops is neglected. The simulation was carried out for 2 million MC cycles for equilibration and another 2 million MC cycles to sample averages. After reaching equilibration, the structural arrangement of particles on the surface is characterized by calculating coordination number distributions. 
\begin{table}[ht]
\caption{Parameters used in the simulation. Non-dimensional parameter $A$ is obtained from Eq. \eqref{eq:parA} with $\lambda_B$ = 0.702 nm, $\sigma$ = 1 $\mu$m, $q_+$ = 2000. The values are given for $R_c$ = 1}
\centering
\begin{tabular}{cc}
\hline
$\phi_+$ &\quad\quad $A\times 10^7$\\ 
\hline
0.45 &\quad\quad 0.98  \\
 0.50 &\quad\quad 1.02  \\
 0.55 &\quad\quad 1.05  \\
 0.6 &\quad\quad 1.09  \\
 0.65&\quad\quad 1.13 \\
 0.70&\quad\quad 1.17 \\
 0.75&\quad\quad 1.22 \\
\hline
\end{tabular}
\end{table}
\subsection{Stability criterion} \label{ssec:criterion}
Interfacial tension and bending modulus of the particle-coated interface are important parameters in the stability of emulsions. However, in some cases, the self-energy of the interface due to inter-particle interactions can be used as a criterion to define stability. Emulsion drops destabilize and lead to phase separation because of coalescence. Pickering emulsion drops are usually stable against coalescence because, as they come rather close to each other, the particles at the interface offer steric hindrance \cite{binks2002}. There are studies which showed that poorly covered emulsions are also stable against coalescence \cite{vignati}. There seems to be a range of coverage depending upon the type of emulsion, type of particles and their interactions that corresponds to stable emulsions. We fix a coverage of 74\% in our studies. Once we fix the required particle coverage, then the question of stability is whether such a coverage is thermodynamically feasible or not. To this end, we can write free energy change due to adsorption as
\begin{equation} \label {eq:freeenergy}
\begin{aligned}
\Delta f = \frac{2\phi S}{N_t}(\gamma_{pw}+\gamma_{po}-\gamma_{ow}) + e_s
\end{aligned}
\end{equation}
Here, $\Delta f$ is the Helmholtz free energy per particle in the units of $k_BT$, $\gamma$ is the interfacial tension, with suffix $p$, $o$ and $w$ denoting particle, oil and water, respectively. The term $e_s$ refers to self-energy of the interface per particle due to inter-particle interactions. In our simulations, we have taken the contact angle of particles as 90$^{\circ}$. From Young-Dupr\' e equation, this means $\gamma_{pw} = \gamma_{po}$. Hence for $\theta = 90^{\circ}$ and $\gamma_{pw} + \gamma_{po} = \gamma_{ow}$, Eq. \eqref{eq:freeenergy} becomes as:
\begin{equation}
\begin{aligned}
\Delta f = e_s
\end{aligned}
\end{equation}
Note that we have neglected the entropy of mixing of particles as it gives a negative contribution in the free energy (discussed later) and that we set an upper bound for stability criterion. Therefore, if the free energy is negative, it is thermodynamically favorable to pack oppositely charged particles at the interface leading to steric stabilization of emulsion drops. In the special case considered here, the interfacial energy change due to adsorption of particle is zero. Therefore, the sign of the interaction energy due to inter-particle interaction then becomes the important factor in determining stability. To be precise, we are considering here a \textit{conditional stability} because of an imposed particle coverage and contact angle. 

\section{Results and Discussion}
\subsection{Emulsion stability}
\begin{figure} 
\centering
\includegraphics [scale=0.33]{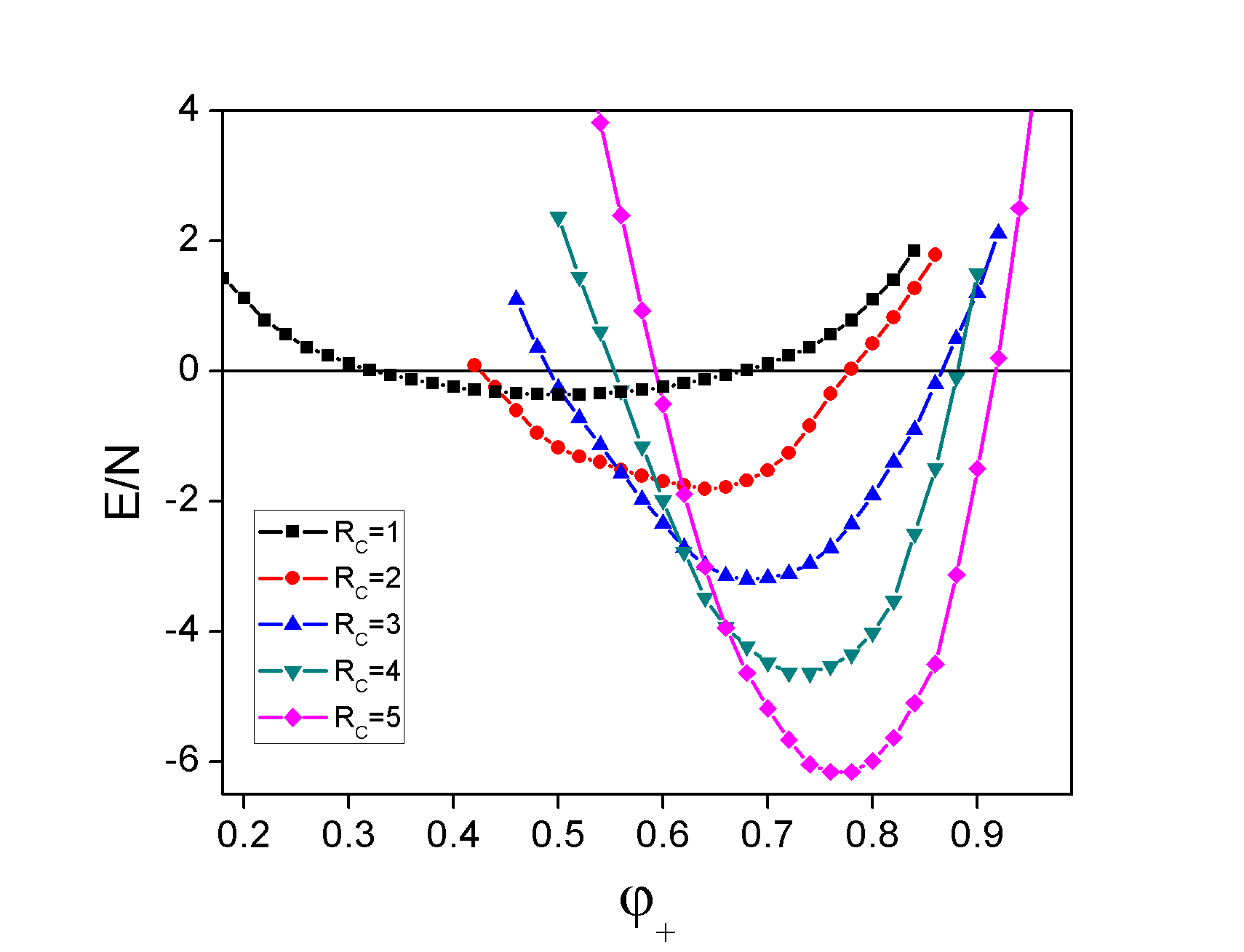}
\caption{(Colour online) Energy per particle as a function of fraction of $\phi_+$ for different ratios of charge density.}
\label {fig:energy}
\end{figure}

\begin{figure} 
\centering
\includegraphics [scale=0.22]{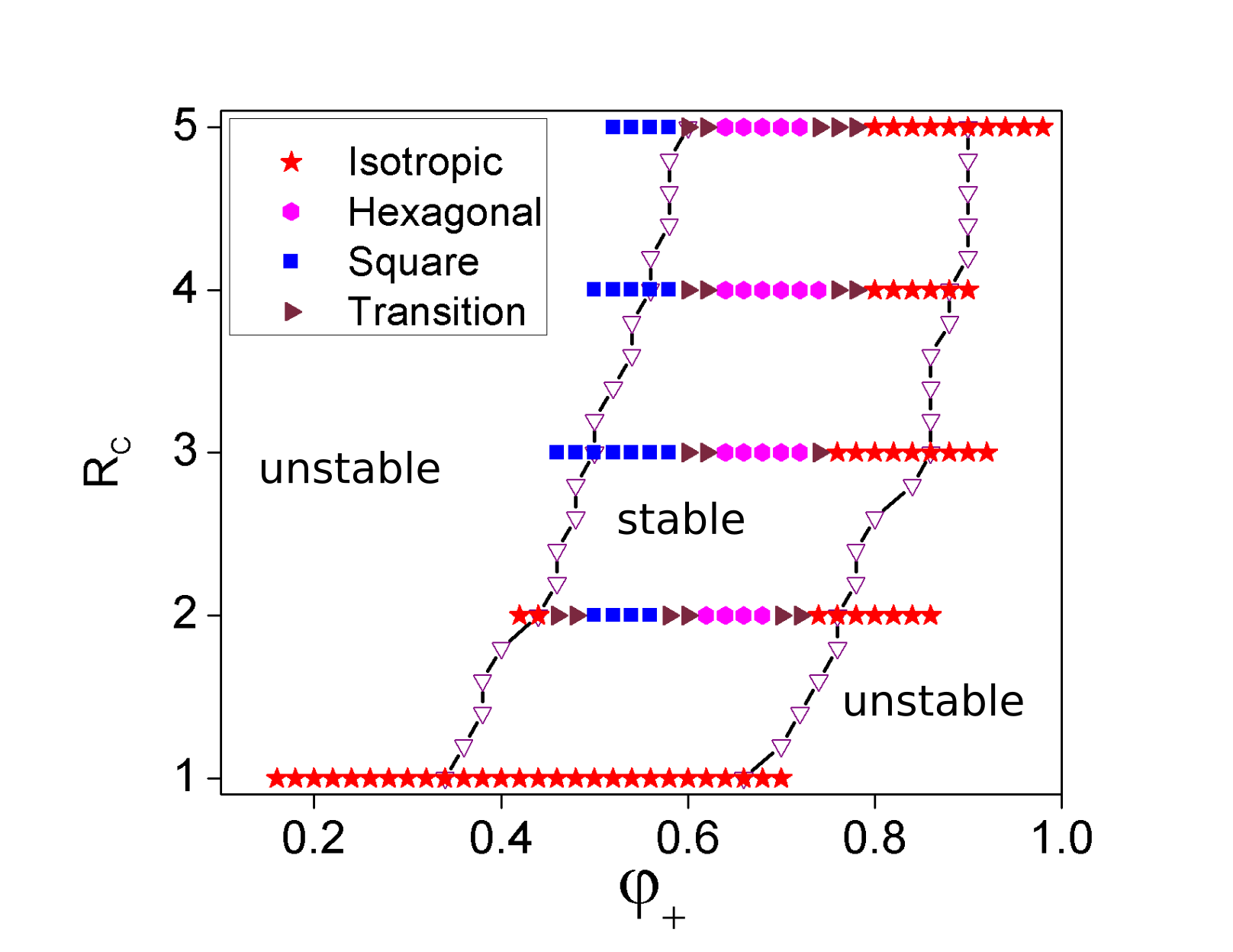}
\caption{(Colour online) Stability diagram of emulsions stabilized by oppositely charged particles based on energy per particle. The connected lines are the boundaries at which energy per particle changes its sign. The markers denote different structural patterns.}
\label {fig:stability}
\end{figure}

We consider emulsion drops with particle-coverage of 74\%, and calculate the energy per particle at equilibrium for several number fractions of oppositely charged particles. Fig. \ref{fig:energy} shows the energy per particle as a function of composition for several charge ratios ($R_c$) at equilibrium. The energy per particle varies steeply with composition as the charge ratio increases from 2 to 5. We see that for every charge ratio, there is a minimum in the energy per particle, leading to an optimum composition. Optimum composition can be correlated to the most stable state of the emulsion drop, leading to better stability. Optimum composition shifts to higher compositions as the charge ratio increases. This is because as the charge ratio is increased, more number of positively charged particles is required to minimize the energy. If an emulsion drop should be stable, then the energy per particle should be zero or negative at a given coverage. According to this definition of stability, we can find that the curves given in Fig. \ref{fig:energy} dictates a range of compositions where emulsions will be stable. This stability region is demarked in Fig. \ref{fig:stability} as the stability diagram. The region between the two connected lines corresponds to stable emulsions. \\

From experimental data, it was found that the stable region for decane/water emulsions stabilized by a mixture of oppositely but equally charged 0.5 $\mu$m polystyrene particles ranges from 0.3 - 0.7 \cite{nallamilli2015}. The simulations shows this range to be 0.34 - 0.66 with a corresponding charge ratio of 1. Despite the model and stability criterion used in our simulations are simplified, the comparison of simulation data with experiments is in reasonable agreement. Note that in our simulations we have considered contact angle of particles as 90$^{\circ}$, and the present simulation results are valid for both o/w and w/o emulsion types. Outside the stable region, it is unlikely that an emulsion would be stable because in these cases energy per particle becomes positive. Therefore, the limit on stability in terms of charge ratio can be useful for rational design of emulsion stabilized by oppositely charged colloids. \\

To verify if entropy of mixing affects the value of optimum composition, we calculate entropy from
\begin{equation}\label{eq:entropy}
\begin{aligned}
S = k_B ln\Omega
\end{aligned}
\end{equation}  
where $\Omega$ is the total number of configurations obtained by mixing $N_+$ positive colloids and $N_-$ negative colloids. Assuming full coverage of the emulsion surface (i.e 90.6\%, although simulations corresponds to a maximum coverage of 74\%), we find that
\begin{equation}\label{eq:omega}
\begin{aligned}
\Omega = \frac{N!}{N_+! N_-!}
\end{aligned}
\end{equation}  
Using Stirling's approximation [$ln(x!) \approx x ln(x)-x$] in Eq.\eqref{eq:omega} and substituting in Eq.\eqref{eq:entropy}, entropy per particle becomes
\begin{equation}
\begin{aligned}
s(\phi_+)=-k_B \phi_+ ln \phi_+ - k_B (1-\phi_+)ln(1-\phi_+)
\end{aligned}
\end{equation}  
Since the actual coverage of particles is 74\%, the excess area available for particles contribute to entropy as $k_Bln(0.906S/0.74S)$, hence the entropy of mixing is given by: 
\begin{equation}
\begin{aligned}
s(\phi_+)=-k_B[\phi_+ln\phi_++(1-\phi_+)ln(1-\phi_+)-ln(0.906/0.74)]
\end{aligned}
\end{equation}  
Then Helmholtz free energy per particle is given by:
\begin{equation}
\begin{aligned}
f(\phi_+) = u(\phi_+) - Ts(\phi_+)
\end{aligned}
\end{equation}  
Note that the change of interfacial energy due to particle adsorption is considered to be zero as discussed earlier in Section \ref{ssec:criterion}

\begin{figure} 
\centering
\includegraphics [scale=0.33]{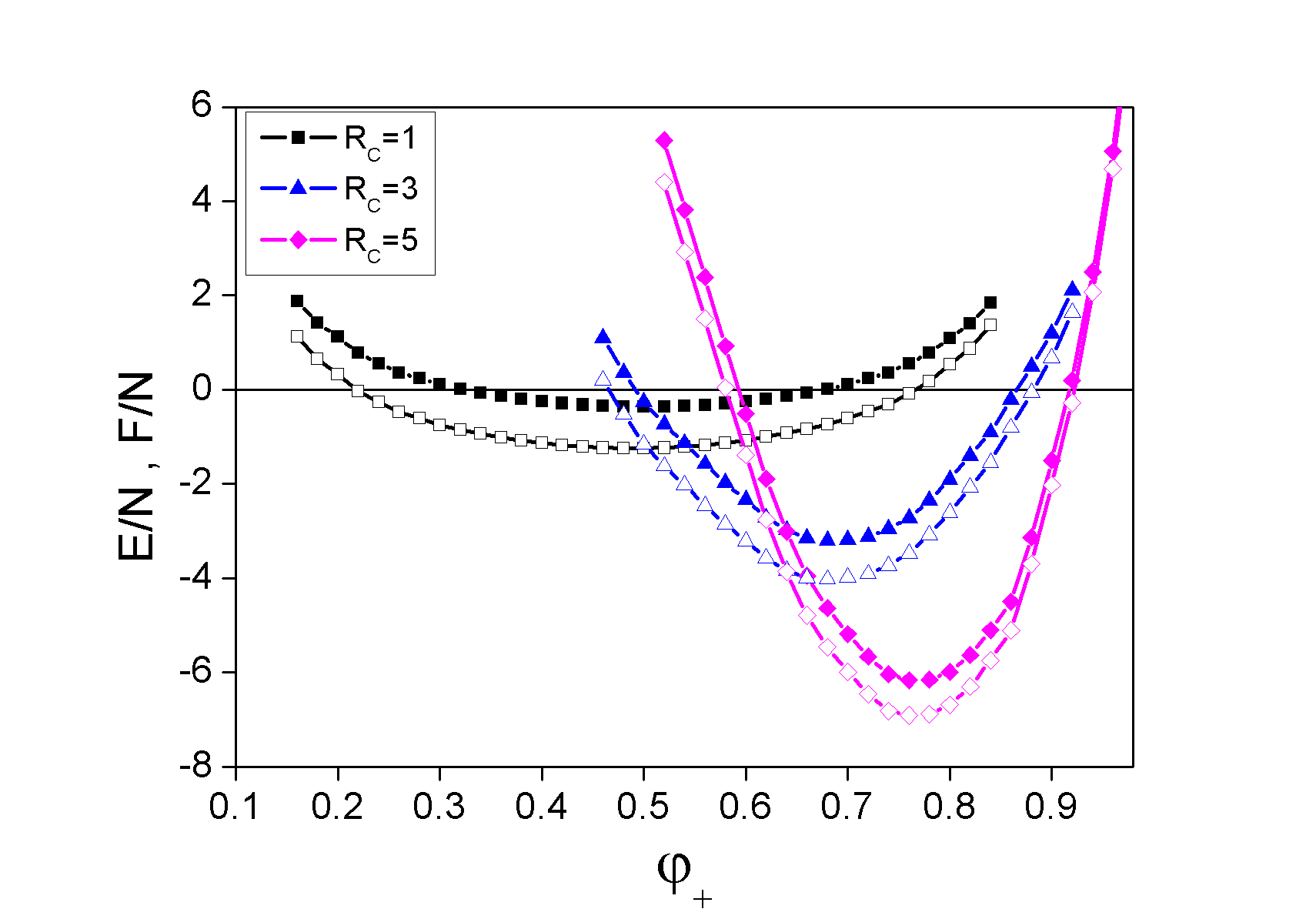}
\caption{(Colour online) Energy (filled markers) and free energy (open markers) per particle as a function of composition $\phi_+$ of for different charge ratio.}
\label {fig:free}
\end{figure}

As shown in Fig. \ref{fig:free}, the effect of mixing entropy is negligible compared to interaction energy in determining the optimum composition for a given charge ratio of colloids. The optimum composition is primarily influenced by dipolar interaction energy. However, the range of stability is slightly affected by entropic contributions. This additional entropic part slightly increases the window of stability.

\subsection{Patterning of oppositely charged colloids at the interface}
\begin{figure}
\begin{center}
\includegraphics[scale=0.4]{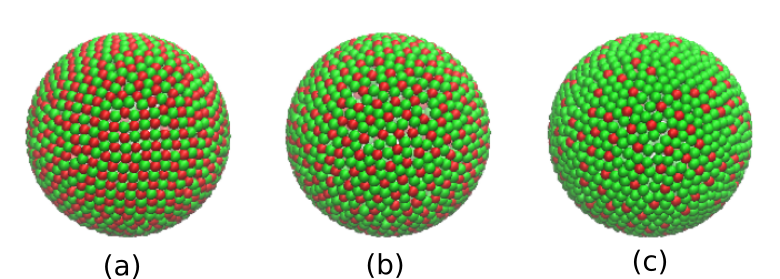}
\caption{(Colour online) Pattern formation on emulsion surface for $R_c$ = 3: a) square patterning for  $\phi_+$ = 0.52, b) hexagonal patterning for $\phi_+$ = 0.64 and c) isotropic phase for $\phi_+$ = 0.82.}
\label {fig:pattern}
\end{center}
\end{figure}

For the imposed coverage, the binary mixture of particles exhibit interesting patterns on the surface of emulsion drop as a function of composition. We classify these patterns into three distinct classes i) isotropic, ii) square and iii) hexagonal patterns. In Fig. \ref{fig:stability}, these patterns are marked for different charge ratios. A snapshot of these phases are shown in Fig. \ref{fig:pattern} for R$_c$ = 3. For R$_c$ = 1, the particles are in completely mixed state, which we call as isotropic phase due to the absence of any structural order in the arrangement of particles. For R$_c>1$, however, we find that all three phases exist. In the square phase, each particle is surrounded by four other type of particles. In the hexagonal phase, minority particles (negatively charged) assemble themselves in a hexagonal lattice, and surrounded by nearly six positively charged particles. If we consider only the stable region, for R$_c$ = 1, all configurations are in isotropic phase. For $2\geq R_c \leq4$, all the three phases are attainable and for R$_c$ = 5, only hexagonal and isotropic phases are attainable. Between the ordered phases, and between ordered and isotropic phases, there are transition phases. For instance, in the transition phase between square and hexagonal phases, both phases coexist on the emulsion surface. Due to the curvature of the interface, the two-dimensional crystal lattices develop many defects and grain boundaries. These imperfections can be minimized if we consider higher coverages close to close-packing limit.\\
 
\begin{figure}
\begin{center}
\includegraphics[scale=0.32]{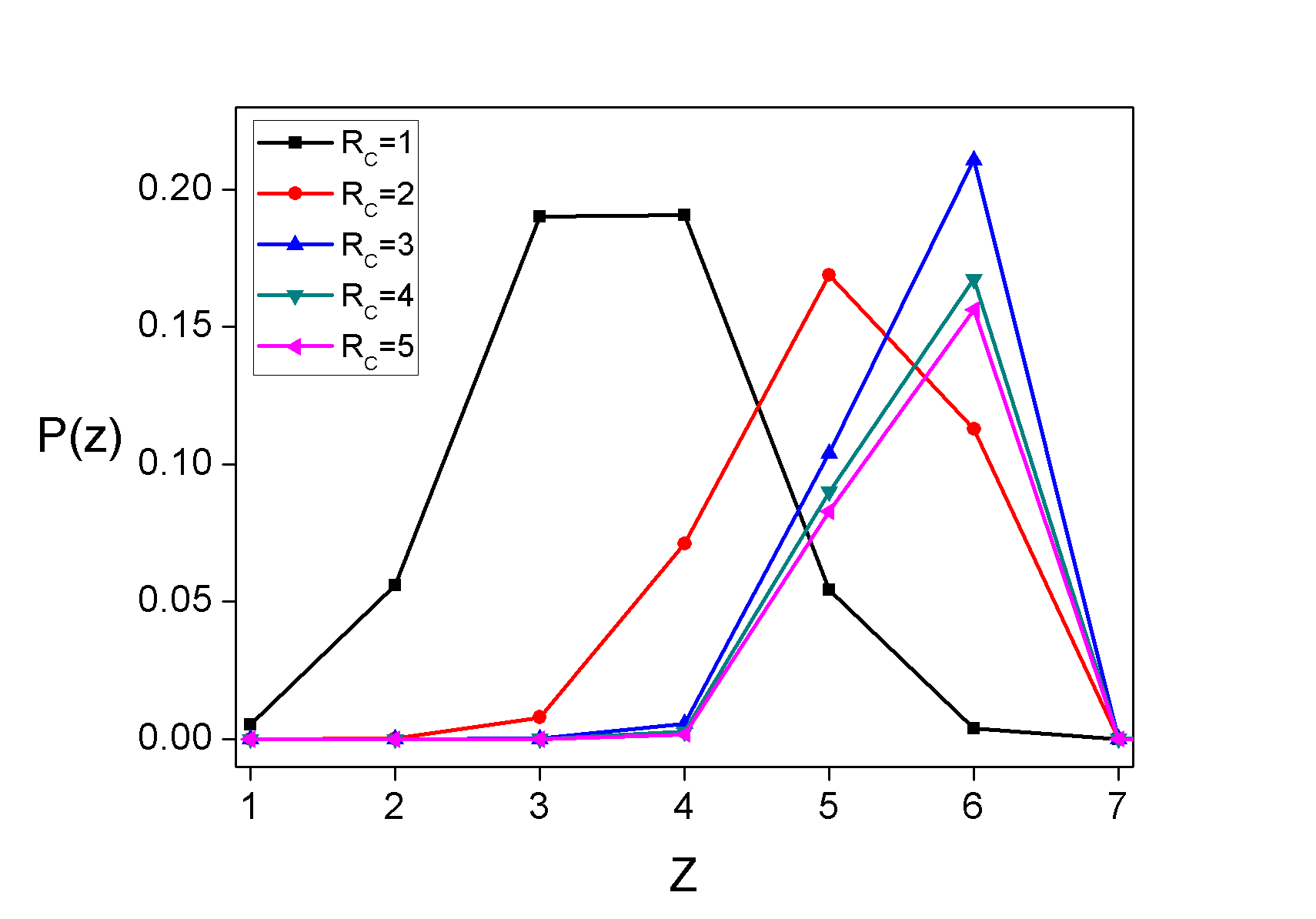}
\caption{(Colour online) Neighbor distribution at optimum compositions for several charge ratios. The compositions correspond to 0.5, 0.64, 0.68, 0.74 and 0.76 for $R_c$ from 1 to 5, respectively.}
\label {fig:neighbor}
\end{center}
\end{figure}
Fig. \ref{fig:neighbor} shows the number of positively charged particles distributed around a negatively charged particle. This data corresponds to the optimum composition for each charge ratio (see Fig. \ref{fig:energy}). The average number of neighbors increases from 3.5 to 5.7 as charge ratio is increased from 1 to 5, although all the simulations correspond to the same coverage. For charge ratio of 3 - 5, the peak at Z = 6 clearly indicates that the particles self-assembly into a hexagonal order. Therefore, for the synthesis of colloidosomes with square or hexagonal pattern, emulsions can be prepared by choosing the right set of experimental conditions (charge ratio and composition) from Fig. \ref{fig:stability}. Note that colloidosome with square lattice can not be made with single type particles, and the use of oppositely charged colloids opens up this possibility. \\

\section{Conclusion}
Use of oppositely charged colloids offers various possibilities to stabilize water in oil or oil in water emulsions by means of controlling the charge ratio and number of particles on their surfaces. From Monte Carlo simulations, we trace a range of composition of particle mixture to stabilize emulsion, and report optimum composition corresponding to most stable state. We find that the charge ratio is an important parameter that controls emulsion stability. Additionally, it is seen that the spatial distribution (patterns) of colloids on the emulsion surface can be modulated with charge density ratio. The simulation study can help experimentalist to design optimum parameters to stabilize emulsions. Further, our approach can in principle be extended to several studies that have shown unusual role of charged colloids at interfaces such as two-dimensional alloys \cite{law}, colloidal gels \cite{sanz}, bridging or bilayer formation at interfaces \cite{horozov2006}. When pickering emulsion stabilized by oppositely charged particles are used to synthesize colloidosomes, then the crystalline phases exhibited at the interface can be used for systematic functionalization and for creating ordered pores.

\acknowledgments
The authors thank Prof. Harmut L\"{o}wen, Prof. Holger Stark and Dr. Adam Law for stimulating discussion. Virgo High-Performance Computing Facility (HPC) at IIT-Madras is acknowledged for the computational resources. EM acknowledges Department of Science and Technology (DST), India for financial support through research grant SR/S3/CE/055/2012.
\bibliographystyle{aipnum4-1}
\bibliography{references}

\begin{thebibliography}{46}%
\makeatletter
\providecommand \@ifxundefined [1]{%
 \@ifx{#1\undefined}
}%
\providecommand \@ifnum [1]{%
 \ifnum #1\expandafter \@firstoftwo
 \else \expandafter \@secondoftwo
 \fi
}%
\providecommand \@ifx [1]{%
 \ifx #1\expandafter \@firstoftwo
 \else \expandafter \@secondoftwo
 \fi
}%
\providecommand \natexlab [1]{#1}%
\providecommand \enquote  [1]{``#1''}%
\providecommand \bibnamefont  [1]{#1}%
\providecommand \bibfnamefont [1]{#1}%
\providecommand \citenamefont [1]{#1}%
\providecommand \href@noop [0]{\@secondoftwo}%
\providecommand \href [0]{\begingroup \@sanitize@url \@href}%
\providecommand \@href[1]{\@@startlink{#1}\@@href}%
\providecommand \@@href[1]{\endgroup#1\@@endlink}%
\providecommand \@sanitize@url [0]{\catcode `\\12\catcode `\$12\catcode
  `\&12\catcode `\#12\catcode `\^12\catcode `\_12\catcode `\%12\relax}%
\providecommand \@@startlink[1]{}%
\providecommand \@@endlink[0]{}%
\providecommand \url  [0]{\begingroup\@sanitize@url \@url }%
\providecommand \@url [1]{\endgroup\@href {#1}{\urlprefix }}%
\providecommand \urlprefix  [0]{URL }%
\providecommand \Eprint [0]{\href }%
\providecommand \doibase [0]{http://dx.doi.org/}%
\providecommand \selectlanguage [0]{\@gobble}%
\providecommand \bibinfo  [0]{\@secondoftwo}%
\providecommand \bibfield  [0]{\@secondoftwo}%
\providecommand \translation [1]{[#1]}%
\providecommand \BibitemOpen [0]{}%
\providecommand \bibitemStop [0]{}%
\providecommand \bibitemNoStop [0]{.\EOS\space}%
\providecommand \EOS [0]{\spacefactor3000\relax}%
\providecommand \BibitemShut  [1]{\csname bibitem#1\endcsname}%
\let\auto@bib@innerbib\@empty
%</preamble>
\bibitem [{\citenamefont {Nallamilli}, \citenamefont {Mani},\ and\
  \citenamefont {Basavaraj}(2014)}]{nallamilli}%
  \BibitemOpen
  \bibfield  {author} {\bibinfo {author} {\bibfnamefont {T.}~\bibnamefont
  {Nallamilli}}, \bibinfo {author} {\bibfnamefont {E.}~\bibnamefont {Mani}}, \
  and\ \bibinfo {author} {\bibfnamefont {M.~G.}\ \bibnamefont {Basavaraj}},\
  }\href {\doibase 10.1021/la501785y} {\bibfield  {journal} {\bibinfo
  {journal} {Langmuir}\ }\textbf {\bibinfo {volume} {30}},\ \bibinfo {pages}
  {9336} (\bibinfo {year} {2014})}\BibitemShut {NoStop}%
\bibitem [{\citenamefont {Nallamilli}\ \emph {et~al.}(2015)\citenamefont
  {Nallamilli}, \citenamefont {Binks}, \citenamefont {Mani},\ and\
  \citenamefont {Basavaraj}}]{nallamilli2015}%
  \BibitemOpen
  \bibfield  {author} {\bibinfo {author} {\bibfnamefont {T.}~\bibnamefont
  {Nallamilli}}, \bibinfo {author} {\bibfnamefont {B.~P.}\ \bibnamefont
  {Binks}}, \bibinfo {author} {\bibfnamefont {E.}~\bibnamefont {Mani}}, \ and\
  \bibinfo {author} {\bibfnamefont {M.~G.}\ \bibnamefont {Basavaraj}},\
  }\href@noop {} {\bibfield  {journal} {\bibinfo  {journal} {Langmuir}\
  }\textbf {\bibinfo {volume} {31}},\ \bibinfo {pages} {11200} (\bibinfo {year}
  {2015})}\BibitemShut {NoStop}%
\bibitem [{\citenamefont {Amalvy}\ \emph {et~al.}(2004)\citenamefont {Amalvy},
  \citenamefont {Unali}, \citenamefont {Li}, \citenamefont {Granger-Bevan},
  \citenamefont {Armes}, \citenamefont {Binks}, \citenamefont {Rodrigues},\
  and\ \citenamefont {Whitby}}]{amalvy}%
  \BibitemOpen
  \bibfield  {author} {\bibinfo {author} {\bibfnamefont {J.~I.}\ \bibnamefont
  {Amalvy}}, \bibinfo {author} {\bibfnamefont {G.-F.}\ \bibnamefont {Unali}},
  \bibinfo {author} {\bibfnamefont {Y.}~\bibnamefont {Li}}, \bibinfo {author}
  {\bibfnamefont {S.}~\bibnamefont {Granger-Bevan}}, \bibinfo {author}
  {\bibfnamefont {S.~P.}\ \bibnamefont {Armes}}, \bibinfo {author}
  {\bibfnamefont {B.~P.}\ \bibnamefont {Binks}}, \bibinfo {author}
  {\bibfnamefont {J.~A.}\ \bibnamefont {Rodrigues}}, \ and\ \bibinfo {author}
  {\bibfnamefont {C.~P.}\ \bibnamefont {Whitby}},\ }\href {\doibase
  10.1021/la035921c} {\bibfield  {journal} {\bibinfo  {journal} {Langmuir}\
  }\textbf {\bibinfo {volume} {20}},\ \bibinfo {pages} {4345} (\bibinfo {year}
  {2004})}\BibitemShut {NoStop}%
\bibitem [{\citenamefont {Lin}\ \emph {et~al.}(2003)\citenamefont {Lin},
  \citenamefont {Skaff}, \citenamefont {Emrick}, \citenamefont {Dinsmore},\
  and\ \citenamefont {Russell}}]{lin}%
  \BibitemOpen
  \bibfield  {author} {\bibinfo {author} {\bibfnamefont {Y.}~\bibnamefont
  {Lin}}, \bibinfo {author} {\bibfnamefont {H.}~\bibnamefont {Skaff}}, \bibinfo
  {author} {\bibfnamefont {T.}~\bibnamefont {Emrick}}, \bibinfo {author}
  {\bibfnamefont {A.~D.}\ \bibnamefont {Dinsmore}}, \ and\ \bibinfo {author}
  {\bibfnamefont {T.~P.}\ \bibnamefont {Russell}},\ }\href {\doibase
  10.1126/science.1078616} {\bibfield  {journal} {\bibinfo  {journal}
  {Science}\ }\textbf {\bibinfo {volume} {299}},\ \bibinfo {pages} {226}
  (\bibinfo {year} {2003})}\BibitemShut {NoStop}%
\bibitem [{\citenamefont {Leunissen}\ \emph {et~al.}(2007)\citenamefont
  {Leunissen}, \citenamefont {van Blaaderen}, \citenamefont {Hollingsworth},
  \citenamefont {Sullivan},\ and\ \citenamefont {Chaikin}}]{leunissen}%
  \BibitemOpen
  \bibfield  {author} {\bibinfo {author} {\bibfnamefont {M.~E.}\ \bibnamefont
  {Leunissen}}, \bibinfo {author} {\bibfnamefont {A.}~\bibnamefont {van
  Blaaderen}}, \bibinfo {author} {\bibfnamefont {A.~D.}\ \bibnamefont
  {Hollingsworth}}, \bibinfo {author} {\bibfnamefont {M.~T.}\ \bibnamefont
  {Sullivan}}, \ and\ \bibinfo {author} {\bibfnamefont {P.~M.}\ \bibnamefont
  {Chaikin}},\ }\href {\doibase 10.1073/pnas.0610589104} {\bibfield  {journal}
  {\bibinfo  {journal} {Proceedings of the National Academy of Sciences}\
  }\textbf {\bibinfo {volume} {104}},\ \bibinfo {pages} {2585} (\bibinfo {year}
  {2007})}\BibitemShut {NoStop}%
\bibitem [{\citenamefont {Bleibel}, \citenamefont {Domínguez},\ and\
  \citenamefont {Oettel}(2013)}]{bleibel}%
  \BibitemOpen
  \bibfield  {author} {\bibinfo {author} {\bibfnamefont {J.}~\bibnamefont
  {Bleibel}}, \bibinfo {author} {\bibfnamefont {A.}~\bibnamefont {Domínguez}},
  \ and\ \bibinfo {author} {\bibfnamefont {M.}~\bibnamefont {Oettel}},\ }\href
  {\doibase 10.1140/epjst/e2013-02076-9} {\bibfield  {journal} {\bibinfo
  {journal} {The European Physical Journal Special Topics}\ }\textbf {\bibinfo
  {volume} {222}},\ \bibinfo {pages} {3071} (\bibinfo {year}
  {2013})}\BibitemShut {NoStop}%
\bibitem [{\citenamefont {Binks}\ and\ \citenamefont {Lumsdon}(2001)}]{binks}%
  \BibitemOpen
  \bibfield  {author} {\bibinfo {author} {\bibfnamefont {B.~P.}\ \bibnamefont
  {Binks}}\ and\ \bibinfo {author} {\bibfnamefont {S.~O.}\ \bibnamefont
  {Lumsdon}},\ }\href {\doibase 10.1021/la0103822} {\bibfield  {journal}
  {\bibinfo  {journal} {Langmuir}\ }\textbf {\bibinfo {volume} {17}},\ \bibinfo
  {pages} {4540} (\bibinfo {year} {2001})}\BibitemShut {NoStop}%
\bibitem [{\citenamefont {Witten}\ and\ \citenamefont {Pincus}(1986)}]{witten}%
  \BibitemOpen
  \bibfield  {author} {\bibinfo {author} {\bibfnamefont {T.~A.}\ \bibnamefont
  {Witten}}\ and\ \bibinfo {author} {\bibfnamefont {P.~A.}\ \bibnamefont
  {Pincus}},\ }\href {\doibase 10.1021/ma00164a009} {\bibfield  {journal}
  {\bibinfo  {journal} {Macromolecules}\ }\textbf {\bibinfo {volume} {19}},\
  \bibinfo {pages} {2509} (\bibinfo {year} {1986})}\BibitemShut {NoStop}%
\bibitem [{\citenamefont {Zhulina}, \citenamefont {Borisov},\ and\
  \citenamefont {Priamitsyn}(1990)}]{zhulina}%
  \BibitemOpen
  \bibfield  {author} {\bibinfo {author} {\bibfnamefont {E.~B.}\ \bibnamefont
  {Zhulina}}, \bibinfo {author} {\bibfnamefont {O.~V.}\ \bibnamefont
  {Borisov}}, \ and\ \bibinfo {author} {\bibfnamefont {V.~A.}\ \bibnamefont
  {Priamitsyn}},\ }\href {\doibase
  http://dx.doi.org/10.1016/0021-9797(90)90423-L} {\bibfield  {journal}
  {\bibinfo  {journal} {Journal of Colloid and Interface Science}\ }\textbf
  {\bibinfo {volume} {137}},\ \bibinfo {pages} {495 } (\bibinfo {year}
  {1990})}\BibitemShut {NoStop}%
\bibitem [{\citenamefont {Sun}\ \emph {et~al.}(2007{\natexlab{a}})\citenamefont
  {Sun}, \citenamefont {Sun}, \citenamefont {Wei}, \citenamefont {Liu},\ and\
  \citenamefont {Zhang}}]{sun}%
  \BibitemOpen
  \bibfield  {author} {\bibinfo {author} {\bibfnamefont {W.}~\bibnamefont
  {Sun}}, \bibinfo {author} {\bibfnamefont {D.}~\bibnamefont {Sun}}, \bibinfo
  {author} {\bibfnamefont {Y.}~\bibnamefont {Wei}}, \bibinfo {author}
  {\bibfnamefont {S.}~\bibnamefont {Liu}}, \ and\ \bibinfo {author}
  {\bibfnamefont {S.}~\bibnamefont {Zhang}},\ }\href {\doibase
  http://dx.doi.org/10.1016/j.jcis.2007.02.082} {\bibfield  {journal} {\bibinfo
   {journal} {Journal of Colloid and Interface Science}\ }\textbf {\bibinfo
  {volume} {311}},\ \bibinfo {pages} {228 } (\bibinfo {year}
  {2007}{\natexlab{a}})}\BibitemShut {NoStop}%
\bibitem [{\citenamefont {Sun}\ \emph {et~al.}(2007{\natexlab{b}})\citenamefont
  {Sun}, \citenamefont {Sun}, \citenamefont {Wei}, \citenamefont {Liu},\ and\
  \citenamefont {Zhang}}]{herzig}%
  \BibitemOpen
  \bibfield  {author} {\bibinfo {author} {\bibfnamefont {W.}~\bibnamefont
  {Sun}}, \bibinfo {author} {\bibfnamefont {D.}~\bibnamefont {Sun}}, \bibinfo
  {author} {\bibfnamefont {Y.}~\bibnamefont {Wei}}, \bibinfo {author}
  {\bibfnamefont {S.}~\bibnamefont {Liu}}, \ and\ \bibinfo {author}
  {\bibfnamefont {S.}~\bibnamefont {Zhang}},\ }\href@noop {} {\bibfield
  {journal} {\bibinfo  {journal} {Nature Materials}\ }\textbf {\bibinfo
  {volume} {6}},\ \bibinfo {pages} {966 } (\bibinfo {year}
  {2007}{\natexlab{b}})}\BibitemShut {NoStop}%
\bibitem [{\citenamefont {Dugyala}, \citenamefont {Daware},\ and\ \citenamefont
  {Basavaraj}(2013)}]{dugyala}%
  \BibitemOpen
  \bibfield  {author} {\bibinfo {author} {\bibfnamefont {V.~R.}\ \bibnamefont
  {Dugyala}}, \bibinfo {author} {\bibfnamefont {S.~V.}\ \bibnamefont {Daware}},
  \ and\ \bibinfo {author} {\bibfnamefont {M.~G.}\ \bibnamefont {Basavaraj}},\
  }\href@noop {} {\bibfield  {journal} {\bibinfo  {journal} {Soft Matter}\
  }\textbf {\bibinfo {volume} {9}},\ \bibinfo {pages} {6711} (\bibinfo {year}
  {2013})}\BibitemShut {NoStop}%
\bibitem [{\citenamefont {Damodaran}(2005)}]{damodaran}%
  \BibitemOpen
  \bibfield  {author} {\bibinfo {author} {\bibfnamefont {S.}~\bibnamefont
  {Damodaran}},\ }\href {\doibase 10.1111/j.1365-2621.2005.tb07150.x}
  {\bibfield  {journal} {\bibinfo  {journal} {Journal of Food Science}\
  }\textbf {\bibinfo {volume} {70}},\ \bibinfo {pages} {R54} (\bibinfo {year}
  {2005})}\BibitemShut {NoStop}%
\bibitem [{\citenamefont {Clark}\ \emph {et~al.}(1994)\citenamefont {Clark},
  \citenamefont {Mackie}, \citenamefont {Wilde},\ and\ \citenamefont
  {Wilson}}]{clark}%
  \BibitemOpen
  \bibfield  {author} {\bibinfo {author} {\bibfnamefont {D.~C.}\ \bibnamefont
  {Clark}}, \bibinfo {author} {\bibfnamefont {A.~R.}\ \bibnamefont {Mackie}},
  \bibinfo {author} {\bibfnamefont {P.~J.}\ \bibnamefont {Wilde}}, \ and\
  \bibinfo {author} {\bibfnamefont {D.~R.}\ \bibnamefont {Wilson}},\ }\href
  {\doibase 10.1039/FD9949800253} {\bibfield  {journal} {\bibinfo  {journal}
  {Faraday Discuss.}\ }\textbf {\bibinfo {volume} {98}},\ \bibinfo {pages}
  {253} (\bibinfo {year} {1994})}\BibitemShut {NoStop}%
\bibitem [{\citenamefont {Walther}, \citenamefont {Hoffmann},\ and\
  \citenamefont {Müller}(2008)}]{walther}%
  \BibitemOpen
  \bibfield  {author} {\bibinfo {author} {\bibfnamefont {A.}~\bibnamefont
  {Walther}}, \bibinfo {author} {\bibfnamefont {M.}~\bibnamefont {Hoffmann}}, \
  and\ \bibinfo {author} {\bibfnamefont {A.}~\bibnamefont {Müller}},\ }\href
  {\doibase 10.1002/anie.200703224} {\bibfield  {journal} {\bibinfo  {journal}
  {Angewandte Chemie International Edition}\ }\textbf {\bibinfo {volume}
  {47}},\ \bibinfo {pages} {711} (\bibinfo {year} {2008})}\BibitemShut
  {NoStop}%
\bibitem [{\citenamefont {Faria}, \citenamefont {Ruiz},\ and\ \citenamefont
  {Resasco}(2010)}]{faria}%
  \BibitemOpen
  \bibfield  {author} {\bibinfo {author} {\bibfnamefont {J.}~\bibnamefont
  {Faria}}, \bibinfo {author} {\bibfnamefont {M.~P.}\ \bibnamefont {Ruiz}}, \
  and\ \bibinfo {author} {\bibfnamefont {D.~E.}\ \bibnamefont {Resasco}},\
  }\href {\doibase 10.1002/adsc.201000479} {\bibfield  {journal} {\bibinfo
  {journal} {Advanced Synthesis Catalysis}\ }\textbf {\bibinfo {volume}
  {352}},\ \bibinfo {pages} {2359} (\bibinfo {year} {2010})}\BibitemShut
  {NoStop}%
\bibitem [{\citenamefont {Pawar}\ and\ \citenamefont
  {Kretzschmar}(2010)}]{pawar}%
  \BibitemOpen
  \bibfield  {author} {\bibinfo {author} {\bibfnamefont {A.~B.}\ \bibnamefont
  {Pawar}}\ and\ \bibinfo {author} {\bibfnamefont {I.}~\bibnamefont
  {Kretzschmar}},\ }\href {\doibase 10.1002/marc.200900614} {\bibfield
  {journal} {\bibinfo  {journal} {Macromolecular Rapid Communications}\
  }\textbf {\bibinfo {volume} {31}},\ \bibinfo {pages} {150} (\bibinfo {year}
  {2010})}\BibitemShut {NoStop}%
\bibitem [{\citenamefont {Lou}\ and\ \citenamefont {Archer}(2008)}]{lou}%
  \BibitemOpen
  \bibfield  {author} {\bibinfo {author} {\bibfnamefont {X.~W.}\ \bibnamefont
  {Lou}}\ and\ \bibinfo {author} {\bibfnamefont {L.~A.}\ \bibnamefont
  {Archer}},\ }\href {\doibase 10.1002/adma.200702379} {\bibfield  {journal}
  {\bibinfo  {journal} {Advanced Materials}\ }\textbf {\bibinfo {volume}
  {20}},\ \bibinfo {pages} {1853} (\bibinfo {year} {2008})}\BibitemShut
  {NoStop}%
\bibitem [{\citenamefont {Dinsmore}\ \emph {et~al.}(2002)\citenamefont
  {Dinsmore}, \citenamefont {Hsu}, \citenamefont {Nikolaides}, \citenamefont
  {Marquez}, \citenamefont {Bausch},\ and\ \citenamefont {Weitz}}]{dinsmore}%
  \BibitemOpen
  \bibfield  {author} {\bibinfo {author} {\bibfnamefont {A.~D.}\ \bibnamefont
  {Dinsmore}}, \bibinfo {author} {\bibfnamefont {M.~F.}\ \bibnamefont {Hsu}},
  \bibinfo {author} {\bibfnamefont {M.~G.}\ \bibnamefont {Nikolaides}},
  \bibinfo {author} {\bibfnamefont {M.}~\bibnamefont {Marquez}}, \bibinfo
  {author} {\bibfnamefont {A.~R.}\ \bibnamefont {Bausch}}, \ and\ \bibinfo
  {author} {\bibfnamefont {D.~A.}\ \bibnamefont {Weitz}},\ }\href {\doibase
  10.1126/science.1074868} {\bibfield  {journal} {\bibinfo  {journal}
  {Science}\ }\textbf {\bibinfo {volume} {298}},\ \bibinfo {pages} {1006}
  (\bibinfo {year} {2002})}\BibitemShut {NoStop}%
\bibitem [{\citenamefont {Easwaramoorthi}\ \emph {et~al.}(2010)\citenamefont
  {Easwaramoorthi}, \citenamefont {Kim}, \citenamefont {Lim}, \citenamefont
  {Song}, \citenamefont {Suh}, \citenamefont {Sessler},\ and\ \citenamefont
  {Kim}}]{easwaramoorthi}%
  \BibitemOpen
  \bibfield  {author} {\bibinfo {author} {\bibfnamefont {S.}~\bibnamefont
  {Easwaramoorthi}}, \bibinfo {author} {\bibfnamefont {P.}~\bibnamefont {Kim}},
  \bibinfo {author} {\bibfnamefont {J.~M.}\ \bibnamefont {Lim}}, \bibinfo
  {author} {\bibfnamefont {S.}~\bibnamefont {Song}}, \bibinfo {author}
  {\bibfnamefont {H.}~\bibnamefont {Suh}}, \bibinfo {author} {\bibfnamefont
  {J.~L.}\ \bibnamefont {Sessler}}, \ and\ \bibinfo {author} {\bibfnamefont
  {D.}~\bibnamefont {Kim}},\ }\href {\doibase 10.1039/C0JM00863J} {\bibfield
  {journal} {\bibinfo  {journal} {J. Mater. Chem.}\ }\textbf {\bibinfo {volume}
  {20}},\ \bibinfo {pages} {9684} (\bibinfo {year} {2010})}\BibitemShut
  {NoStop}%
\bibitem [{\citenamefont {Liu}\ \emph {et~al.}(2010)\citenamefont {Liu},
  \citenamefont {Liu}, \citenamefont {Dong}, \citenamefont {Yang},\ and\
  \citenamefont {Sun}}]{liu}%
  \BibitemOpen
  \bibfield  {author} {\bibinfo {author} {\bibfnamefont {G.}~\bibnamefont
  {Liu}}, \bibinfo {author} {\bibfnamefont {S.}~\bibnamefont {Liu}}, \bibinfo
  {author} {\bibfnamefont {X.}~\bibnamefont {Dong}}, \bibinfo {author}
  {\bibfnamefont {F.}~\bibnamefont {Yang}}, \ and\ \bibinfo {author}
  {\bibfnamefont {D.}~\bibnamefont {Sun}},\ }\href {\doibase
  http://dx.doi.org/10.1016/j.jcis.2009.08.042} {\bibfield  {journal} {\bibinfo
   {journal} {Journal of Colloid and Interface Science}\ }\textbf {\bibinfo
  {volume} {345}},\ \bibinfo {pages} {302 } (\bibinfo {year}
  {2010})}\BibitemShut {NoStop}%
\bibitem [{\citenamefont {Binks}, \citenamefont {Liu},\ and\ \citenamefont
  {Rodrigues}(2008)}]{binks2008}%
  \BibitemOpen
  \bibfield  {author} {\bibinfo {author} {\bibfnamefont {B.~P.}\ \bibnamefont
  {Binks}}, \bibinfo {author} {\bibfnamefont {W.}~\bibnamefont {Liu}}, \ and\
  \bibinfo {author} {\bibfnamefont {J.~A.}\ \bibnamefont {Rodrigues}},\
  }\href@noop {} {\bibfield  {journal} {\bibinfo  {journal} {Langmuir}\
  }\textbf {\bibinfo {volume} {24}},\ \bibinfo {pages} {4443} (\bibinfo {year}
  {2008})}\BibitemShut {NoStop}%
\bibitem [{\citenamefont {Whitesides}\ and\ \citenamefont
  {Ross}(1995)}]{whitesides}%
  \BibitemOpen
  \bibfield  {author} {\bibinfo {author} {\bibfnamefont {T.~H.}\ \bibnamefont
  {Whitesides}}\ and\ \bibinfo {author} {\bibfnamefont {D.~S.}\ \bibnamefont
  {Ross}},\ }\href {\doibase http://dx.doi.org/10.1006/jcis.1995.1005}
  {\bibfield  {journal} {\bibinfo  {journal} {Journal of Colloid and Interface
  Science}\ }\textbf {\bibinfo {volume} {169}},\ \bibinfo {pages} {48 }
  (\bibinfo {year} {1995})}\BibitemShut {NoStop}%
\bibitem [{\citenamefont {Arditty}\ \emph {et~al.}(2003)\citenamefont
  {Arditty}, \citenamefont {Whitby}, \citenamefont {Binks}, \citenamefont
  {Schmitt},\ and\ \citenamefont {Leal-Calderon}}]{arditty}%
  \BibitemOpen
  \bibfield  {author} {\bibinfo {author} {\bibfnamefont {S.}~\bibnamefont
  {Arditty}}, \bibinfo {author} {\bibfnamefont {C.}~\bibnamefont {Whitby}},
  \bibinfo {author} {\bibfnamefont {B.}~\bibnamefont {Binks}}, \bibinfo
  {author} {\bibfnamefont {V.}~\bibnamefont {Schmitt}}, \ and\ \bibinfo
  {author} {\bibfnamefont {F.}~\bibnamefont {Leal-Calderon}},\ }\href {\doibase
  10.1140/epje/i2003-10018-6} {\bibfield  {journal} {\bibinfo  {journal} {The
  European Physical Journal E}\ }\textbf {\bibinfo {volume} {11}},\ \bibinfo
  {pages} {273} (\bibinfo {year} {2003})}\BibitemShut {NoStop}%
\bibitem [{\citenamefont {Fantoni}, \citenamefont {Salari},\ and\ \citenamefont
  {Klumperman}(2012)}]{fantoni}%
  \BibitemOpen
  \bibfield  {author} {\bibinfo {author} {\bibfnamefont {R.}~\bibnamefont
  {Fantoni}}, \bibinfo {author} {\bibfnamefont {J.~W.~O.}\ \bibnamefont
  {Salari}}, \ and\ \bibinfo {author} {\bibfnamefont {B.}~\bibnamefont
  {Klumperman}},\ }\href {\doibase 10.1103/PhysRevE.85.061404} {\bibfield
  {journal} {\bibinfo  {journal} {Phys. Rev. E}\ }\textbf {\bibinfo {volume}
  {85}},\ \bibinfo {pages} {061404} (\bibinfo {year} {2012})}\BibitemShut
  {NoStop}%
\bibitem [{\citenamefont {Loverde}, \citenamefont {Velichko},\ and\
  \citenamefont {Olvera de~la Cruz}(2006)}]{loverde2006}%
  \BibitemOpen
  \bibfield  {author} {\bibinfo {author} {\bibfnamefont {S.~M.}\ \bibnamefont
  {Loverde}}, \bibinfo {author} {\bibfnamefont {Y.~S.}\ \bibnamefont
  {Velichko}}, \ and\ \bibinfo {author} {\bibfnamefont {M.}~\bibnamefont
  {Olvera de~la Cruz}},\ }\href {\doibase http://dx.doi.org/10.1063/1.2181573}
  {\bibfield  {journal} {\bibinfo  {journal} {The Journal of Chemical Physics}\
  }\textbf {\bibinfo {volume} {124}},\ \bibinfo {eid} {144702} (\bibinfo {year}
  {2006})}\BibitemShut {NoStop}%
\bibitem [{\citenamefont {Loverde}\ and\ \citenamefont {Olvera de~la
  Cruz}(2007)}]{loverde2007}%
  \BibitemOpen
  \bibfield  {author} {\bibinfo {author} {\bibfnamefont {S.~M.}\ \bibnamefont
  {Loverde}}\ and\ \bibinfo {author} {\bibfnamefont {M.}~\bibnamefont {Olvera
  de~la Cruz}},\ }\href {\doibase http://dx.doi.org/10.1063/1.2793038}
  {\bibfield  {journal} {\bibinfo  {journal} {The Journal of Chemical Physics}\
  }\textbf {\bibinfo {volume} {127}},\ \bibinfo {eid} {164707} (\bibinfo {year}
  {2007})}\BibitemShut {NoStop}%
\bibitem [{\citenamefont {Miracle}(2004)}]{miracle}%
  \BibitemOpen
  \bibfield  {author} {\bibinfo {author} {\bibfnamefont {D.~B.}\ \bibnamefont
  {Miracle}},\ }\href@noop {} {\bibfield  {journal} {\bibinfo  {journal}
  {Nature Materials}\ }\textbf {\bibinfo {volume} {3}},\ \bibinfo {pages} {697}
  (\bibinfo {year} {2004})}\BibitemShut {NoStop}%
\bibitem [{\citenamefont {Ohkubo}\ and\ \citenamefont
  {Hirotsu}(2003)}]{ohkubo}%
  \BibitemOpen
  \bibfield  {author} {\bibinfo {author} {\bibfnamefont {T.}~\bibnamefont
  {Ohkubo}}\ and\ \bibinfo {author} {\bibfnamefont {Y.}~\bibnamefont
  {Hirotsu}},\ }\href@noop {} {\bibfield  {journal} {\bibinfo  {journal} {Phys.
  Rev. B}\ }\textbf {\bibinfo {volume} {67}},\ \bibinfo {pages} {094201}
  (\bibinfo {year} {2003})}\BibitemShut {NoStop}%
\bibitem [{\citenamefont {Miracle}\ and\ \citenamefont
  {Senkov}(2003)}]{miracle2003}%
  \BibitemOpen
  \bibfield  {author} {\bibinfo {author} {\bibfnamefont {D.}~\bibnamefont
  {Miracle}}\ and\ \bibinfo {author} {\bibfnamefont {O.}~\bibnamefont
  {Senkov}},\ }\href {\doibase http://dx.doi.org/10.1016/S0022-3093(02)01917-8}
  {\bibfield  {journal} {\bibinfo  {journal} {Journal of Non-Crystalline
  Solids}\ }\textbf {\bibinfo {volume} {319}},\ \bibinfo {pages} {174 }
  (\bibinfo {year} {2003})}\BibitemShut {NoStop}%
\bibitem [{\citenamefont {Sadoc}\ and\ \citenamefont {Dixmier}(1976)}]{sadoc}%
  \BibitemOpen
  \bibfield  {author} {\bibinfo {author} {\bibfnamefont {J.}~\bibnamefont
  {Sadoc}}\ and\ \bibinfo {author} {\bibfnamefont {J.}~\bibnamefont
  {Dixmier}},\ }\href {\doibase http://dx.doi.org/10.1016/0025-5416(76)90192-0}
  {\bibfield  {journal} {\bibinfo  {journal} {Materials Science and
  Engineering}\ }\textbf {\bibinfo {volume} {23}},\ \bibinfo {pages} {187 }
  (\bibinfo {year} {1976})}\BibitemShut {NoStop}%
\bibitem [{\citenamefont {Sadoc}, \citenamefont {Dixmier},\ and\ \citenamefont
  {Guinier}(1973)}]{sadoc1973}%
  \BibitemOpen
  \bibfield  {author} {\bibinfo {author} {\bibfnamefont {J.}~\bibnamefont
  {Sadoc}}, \bibinfo {author} {\bibfnamefont {J.}~\bibnamefont {Dixmier}}, \
  and\ \bibinfo {author} {\bibfnamefont {A.}~\bibnamefont {Guinier}},\ }\href
  {\doibase http://dx.doi.org/10.1016/0022-3093(73)90054-9} {\bibfield
  {journal} {\bibinfo  {journal} {Journal of Non-Crystalline Solids}\ }\textbf
  {\bibinfo {volume} {12}},\ \bibinfo {pages} {46 } (\bibinfo {year}
  {1973})}\BibitemShut {NoStop}%
\bibitem [{\citenamefont {Polk}(1973)}]{polk}%
  \BibitemOpen
  \bibfield  {author} {\bibinfo {author} {\bibfnamefont {D.}~\bibnamefont
  {Polk}},\ }\href {\doibase http://dx.doi.org/10.1016/0022-3093(73)90026-4}
  {\bibfield  {journal} {\bibinfo  {journal} {Journal of Non-Crystalline
  Solids}\ }\textbf {\bibinfo {volume} {11}},\ \bibinfo {pages} {381 }
  (\bibinfo {year} {1973})}\BibitemShut {NoStop}%
\bibitem [{\citenamefont {Assoud}, \citenamefont {Messina},\ and\ \citenamefont
  {L\"{o}wen}(2010)}]{assoud2010}%
  \BibitemOpen
  \bibfield  {author} {\bibinfo {author} {\bibfnamefont {L.}~\bibnamefont
  {Assoud}}, \bibinfo {author} {\bibfnamefont {R.}~\bibnamefont {Messina}}, \
  and\ \bibinfo {author} {\bibfnamefont {H.}~\bibnamefont {L\"{o}wen}},\ }\href
  {http://stacks.iop.org/0295-5075/89/i=3/a=36001} {\bibfield  {journal}
  {\bibinfo  {journal} {EPL (Europhysics Letters)}\ }\textbf {\bibinfo {volume}
  {89}},\ \bibinfo {pages} {36001} (\bibinfo {year} {2010})}\BibitemShut
  {NoStop}%
\bibitem [{\citenamefont {Assoud}, \citenamefont {Messina},\ and\ \citenamefont
  {L\"{o}wen}(2007)}]{assoud}%
  \BibitemOpen
  \bibfield  {author} {\bibinfo {author} {\bibfnamefont {L.}~\bibnamefont
  {Assoud}}, \bibinfo {author} {\bibfnamefont {R.}~\bibnamefont {Messina}}, \
  and\ \bibinfo {author} {\bibfnamefont {H.}~\bibnamefont {L\"{o}wen}},\ }\href
  {http://stacks.iop.org/0295-5075/80/i=4/a=48001} {\bibfield  {journal}
  {\bibinfo  {journal} {EPL (Europhysics Letters)}\ }\textbf {\bibinfo {volume}
  {80}},\ \bibinfo {pages} {48001} (\bibinfo {year} {2007})}\BibitemShut
  {NoStop}%
\bibitem [{\citenamefont {Mani}\ \emph {et~al.}(2010)\citenamefont {Mani},
  \citenamefont {Sanz}, \citenamefont {Bolhuis},\ and\ \citenamefont
  {Kegel}}]{mani2010}%
  \BibitemOpen
  \bibfield  {author} {\bibinfo {author} {\bibfnamefont {E.}~\bibnamefont
  {Mani}}, \bibinfo {author} {\bibfnamefont {E.}~\bibnamefont {Sanz}}, \bibinfo
  {author} {\bibfnamefont {P.~G.}\ \bibnamefont {Bolhuis}}, \ and\ \bibinfo
  {author} {\bibfnamefont {W.~K.}\ \bibnamefont {Kegel}},\ }\href {\doibase
  10.1021/jp1004067} {\bibfield  {journal} {\bibinfo  {journal} {The Journal of
  Physical Chemistry C}\ }\textbf {\bibinfo {volume} {114}},\ \bibinfo {pages}
  {7780} (\bibinfo {year} {2010})}\BibitemShut {NoStop}%
\bibitem [{\citenamefont {Pieranski}(1980)}]{pieranski}%
  \BibitemOpen
  \bibfield  {author} {\bibinfo {author} {\bibfnamefont {P.}~\bibnamefont
  {Pieranski}},\ }\href {\doibase 10.1103/PhysRevLett.45.569} {\bibfield
  {journal} {\bibinfo  {journal} {Phys. Rev. Lett.}\ }\textbf {\bibinfo
  {volume} {45}},\ \bibinfo {pages} {569} (\bibinfo {year} {1980})}\BibitemShut
  {NoStop}%
\bibitem [{\citenamefont {Horozov}\ \emph {et~al.}(2005)\citenamefont
  {Horozov}, \citenamefont {Aveyard}, \citenamefont {Binks},\ and\
  \citenamefont {Clint}}]{horozov}%
  \BibitemOpen
  \bibfield  {author} {\bibinfo {author} {\bibfnamefont {T.~S.}\ \bibnamefont
  {Horozov}}, \bibinfo {author} {\bibfnamefont {R.}~\bibnamefont {Aveyard}},
  \bibinfo {author} {\bibfnamefont {B.~P.}\ \bibnamefont {Binks}}, \ and\
  \bibinfo {author} {\bibfnamefont {J.~H.}\ \bibnamefont {Clint}},\ }\href@noop
  {} {\bibfield  {journal} {\bibinfo  {journal} {Langmuir}\ }\textbf {\bibinfo
  {volume} {21}},\ \bibinfo {pages} {7405} (\bibinfo {year}
  {2005})}\BibitemShut {NoStop}%
\bibitem [{\citenamefont {Law}, \citenamefont {Buzza},\ and\ \citenamefont
  {Horozov}(2011)}]{law}%
  \BibitemOpen
  \bibfield  {author} {\bibinfo {author} {\bibfnamefont {A.~D.}\ \bibnamefont
  {Law}}, \bibinfo {author} {\bibfnamefont {D.~M.~A.}\ \bibnamefont {Buzza}}, \
  and\ \bibinfo {author} {\bibfnamefont {T.~S.}\ \bibnamefont {Horozov}},\
  }\href@noop {} {\bibfield  {journal} {\bibinfo  {journal} {Phys. Rev. Lett.}\
  }\textbf {\bibinfo {volume} {106}},\ \bibinfo {pages} {128302} (\bibinfo
  {year} {2011})}\BibitemShut {NoStop}%
\bibitem [{\citenamefont {Frydel}, \citenamefont {Dietrich},\ and\
  \citenamefont {Oettel}(2007)}]{frydel}%
  \BibitemOpen
  \bibfield  {author} {\bibinfo {author} {\bibfnamefont {D.}~\bibnamefont
  {Frydel}}, \bibinfo {author} {\bibfnamefont {S.}~\bibnamefont {Dietrich}}, \
  and\ \bibinfo {author} {\bibfnamefont {M.}~\bibnamefont {Oettel}},\
  }\href@noop {} {\bibfield  {journal} {\bibinfo  {journal} {Phys. Rev. Lett.}\
  }\textbf {\bibinfo {volume} {99}},\ \bibinfo {pages} {118302} (\bibinfo
  {year} {2007})}\BibitemShut {NoStop}%
\bibitem [{\citenamefont {Saff}\ and\ \citenamefont {Kuijlaars}(1997)}]{saff}%
  \BibitemOpen
  \bibfield  {author} {\bibinfo {author} {\bibfnamefont {E.}~\bibnamefont
  {Saff}}\ and\ \bibinfo {author} {\bibfnamefont {A.}~\bibnamefont
  {Kuijlaars}},\ }\href {\doibase 10.1007/BF03024331} {\bibfield  {journal}
  {\bibinfo  {journal} {The Mathematical Intelligencer}\ }\textbf {\bibinfo
  {volume} {19}},\ \bibinfo {pages} {5} (\bibinfo {year} {1997})}\BibitemShut
  {NoStop}%
\bibitem [{\citenamefont {Tamayo}\ and\ \citenamefont {Klein}(1989)}]{tamayo}%
  \BibitemOpen
  \bibfield  {author} {\bibinfo {author} {\bibfnamefont {P.}~\bibnamefont
  {Tamayo}}\ and\ \bibinfo {author} {\bibfnamefont {W.}~\bibnamefont {Klein}},\
  }\href {\doibase 10.1103/PhysRevLett.63.2757} {\bibfield  {journal} {\bibinfo
   {journal} {Phys. Rev. Lett.}\ }\textbf {\bibinfo {volume} {63}},\ \bibinfo
  {pages} {2757} (\bibinfo {year} {1989})}\BibitemShut {NoStop}%
\bibitem [{\citenamefont {Binks}(2002)}]{binks2002}%
  \BibitemOpen
  \bibfield  {author} {\bibinfo {author} {\bibfnamefont {B.~P.}\ \bibnamefont
  {Binks}},\ }\href@noop {} {\bibfield  {journal} {\bibinfo  {journal} {Current
  Opinion in Colloid and Interface Science}\ }\textbf {\bibinfo {volume} {7}},\
  \bibinfo {pages} {21 } (\bibinfo {year} {2002})}\BibitemShut {NoStop}%
\bibitem [{\citenamefont {Vignati}, \citenamefont {Piazza},\ and\ \citenamefont
  {Lockhart}(2003)}]{vignati}%
  \BibitemOpen
  \bibfield  {author} {\bibinfo {author} {\bibfnamefont {E.}~\bibnamefont
  {Vignati}}, \bibinfo {author} {\bibfnamefont {R.}~\bibnamefont {Piazza}}, \
  and\ \bibinfo {author} {\bibfnamefont {T.~P.}\ \bibnamefont {Lockhart}},\
  }\href {\doibase 10.1021/la034264l} {\bibfield  {journal} {\bibinfo
  {journal} {Langmuir}\ }\textbf {\bibinfo {volume} {19}},\ \bibinfo {pages}
  {6650} (\bibinfo {year} {2003})}\BibitemShut {NoStop}%
\bibitem [{\citenamefont {Sanz}\ \emph {et~al.}(2009)\citenamefont {Sanz},
  \citenamefont {White}, \citenamefont {Clegg}, ,\ and\ \citenamefont
  {Cates}}]{sanz}%
  \BibitemOpen
  \bibfield  {author} {\bibinfo {author} {\bibfnamefont {E.}~\bibnamefont
  {Sanz}}, \bibinfo {author} {\bibfnamefont {K.~A.}\ \bibnamefont {White}},
  \bibinfo {author} {\bibfnamefont {P.~S.}\ \bibnamefont {Clegg}}, , \ and\
  \bibinfo {author} {\bibfnamefont {M.~E.}\ \bibnamefont {Cates}},\ }\href@noop
  {} {\bibfield  {journal} {\bibinfo  {journal} {Phys. Rev. Lett.}\ }\textbf
  {\bibinfo {volume} {103}},\ \bibinfo {pages} {1} (\bibinfo {year}
  {2009})}\BibitemShut {NoStop}%
\bibitem [{\citenamefont {Horozov}\ and\ \citenamefont
  {Binks}(2006)}]{horozov2006}%
  \BibitemOpen
  \bibfield  {author} {\bibinfo {author} {\bibfnamefont {T.~S.}\ \bibnamefont
  {Horozov}}\ and\ \bibinfo {author} {\bibfnamefont {B.~P.}\ \bibnamefont
  {Binks}},\ }\href {\doibase 10.1002/anie.200503131} {\bibfield  {journal}
  {\bibinfo  {journal} {Angewandte Chemie International Edition}\ }\textbf
  {\bibinfo {volume} {45}},\ \bibinfo {pages} {773} (\bibinfo {year}
  {2006})}\BibitemShut {NoStop}%
\end{thebibliography}%
\end{document}